# Comment on "A zero-thermal-quenching phosphor"

Shirun Yan

Department of Chemistry, Fudan University,

Songhu Road 2005, Shanghai 200438,P.R. China

E-mail: sryan@fudan.edu.cn

Kim et al. in a recent article reported a phosphor $Na_{3-2x}Sc_2(PO_4)_3$:$xEu^{2+}$ (NSPO:$xEu^{2+}$) ($\lambda_{em}$ =453 nm) that did not exhibit thermal quenching (TQ) even up to 200 °C[1]. The authors suggested that zero-TQ originates from the compensation of emission losses due to the polymorphic nature of the host and the energy transfer from traps consisting of electron–hole pairs to the 5d-band of $Eu^{2+}$, leading to radiative recombination[1]. The temperature-dependent excitation and emission spectra seemingly supported the authors' assertion. However, a close inspection of the CIE chromaticity coordinates and correlated color temperature (CCT) the WLED prototype fabricated using NSPO:0.03$Eu^{2+}$, $La_3Si_6N_{11}$:$Ce^{3+}$ and (Sr,Ca)$AlSiN_3$:$Eu^{2+}$ as a blue, yellow and red-emitting component respectively (abbreviated as WLED device hereafter) under operating currents of 100-1000 mA alongside a series of optical characterization results of NSPO:$xEu^{2+}$ indicates that the so-called zero-TQ should not be the case. There were some questionable points in optical characterization of the phosphors. The observed increase in intensity upon elevating temperature in the temperature-dependent excitation and emission spectra was likely a pitfall caused by optical path length variation of the spectrofluorimeter induced by lattice thermal expansion and/or phase transitions.

The authors told a good story about zero-TQ and proposed seemingly plausible self-healing mechanism for NSPO:$xEu^{2+}$ phosphors[1]. Nonetheless, CCT of the WLED device under driven currents of 100 - 1000 mA calculated from the CIE chromaticity coordinates x and y in Supplementary Table 5[1] according to McCamy's formula [2] initially shifted to higher values with increasing current from 100 to 300 mA and then shifted steadily to lower values with continuously increasing current above 400 mA, as displayed in Table 1, indicating that the blue phosphor was experiencing luminescence quenching more seriously than the red phosphor above 400 mA. Otherwise, if the blue phosphor did not quench or quenched at the same rate as that of the red phosphor, the CCT of the WLED device should shift to higher values or remain constant. Generally, two issues, i.e. optical saturation and TQ limit the maximum attainable radiance of luminescent materials under high excitation density of the incident light (droop in phosphors)[3-5]. The optical saturation is related to ground-state depletion (thus lowering the absorption strength) or to energy transfer between dopant ions in the excited state[5]. This means that the output luminous flux will no longer increase linearly with the incident power light. The degree of optical saturation is dependent on the relative magnitude of the rates of excitation and de-excitation of the excited state. Phosphors with longer lifetimes are more easily subject to optical saturation than those with shorter decay times[5]. The lifetime of $Eu^{2+}$ luminescence in NSPO:0.03$Eu^{2+}$ (0.39 μs[1]) was shorter than that in $CaAlSiN_3$:$Eu^{2+}$ (0.75μs[6], 0.80μs[7]), hence optical saturation was unlikely the reason for decreasing CCT of the WLED device under high driven currents[5]. Instead, it should originate from TQ. The CCT of the WLED device shifted to low values steadily with increasing current above 400 mA, indicating that TQ of blue-emitting phosphor NSPO:0.03$Eu^{2+}$ was more seriously than that of red-emitting phosphor (Sr,Ca)$AlSiN_3$:$Eu^{2+}$. Zero-TQ for NSPO:0.03$Eu^{2+}$ could not hold unambiguously.

There was no direct experimental proof supporting the existence of energy transfer from traps to $Eu^{2+}$ in NSPO:x$Eu^{2+}$ phosphors. The activation energy of two traps in the β phase NSPO:x$Eu^{2+}$ derived from the TL spectra was 0.75 eV and 0.80 eV respectively[1] which is comparable to the corresponding value in $Sr_4Al_{14}O_{25}:Eu^{2+},Dy^{3+}$ (0.91eV) [8] and $Sr_2MgSi_2O_7:Eu^{2+},Dy^{3+}$ (0.75eV) [9] phosphors. The energy trapped by shallow defects during excitation of $Sr_4Al_{14}O_{25}:Eu^{2+},Dy^{3+}$ or $Sr_2MgSi_2O_7:Eu^{2+},Dy^{3+}$ phosphor could be de-trapped through quantum tunneling or thermal excitation and subsequently transferred to the $Eu^{2+}$, resulting in persistent luminescence of $Eu^{2+}$ in these two phosphors[8,9]. However, no persistent luminescence was observed at either room temperature (RT) or elevated temperatures of 100 and 150 °C. Given that the traps responsible for the zero-TQ property not only exist in a broad band with energy range of 3.1–3.5 eV but also persist in a wide temperature window of 25 – 250 °C under both heating and cooling process as the authors claimed[1], persistent luminescence was expected to occur in NSPO:x$Eu^{2+}$ if the energy transfer from the traps to $Eu^{2+}$ had occurred. Since the whole process including energy storage by traps, de-trapping of the stored energy via thermal stimulation, and energy transfer from the traps to $Eu^{2+}$ and eventually light emission by $Eu^{2+}$ was a rather slower process as compared to direct decay from the 5d excited states to 4f ground state of $Eu^{2+}$. A slight increase in luminescence lifetime upon increasing temperature was probably caused by self-absorption. Fig.2 (a) showed that the phosphor had overlapping emission and absorption bands[1]. The spectra overlap is expected to scale up due to broadening of PL and PLE at elevated temperatures[10]. These overlapping bands inevitably led to self-absorption effects[11], which would give rise to a longer decay time of the emission [12].

Based on the above reasons, one can hardly conclude that zero-TQ by energy transfer from traps to $Eu^{2+}$ for NSPO:x$Eu^{2+}$ phosphors was the case. The observed increasing intensity with increasing temperature, in my opinion, was likely a pitfall caused by shortened optical path lengths between excitation source and phosphor sample as well as between phosphor sample and detector of the measuring spectrofluorimeter induced by lattice expansion and/or phase transitions upon heating during measurements, which subsequently influenced the photoelectric signal recorded by the spectrofluorimeter. The in-situ XRD demonstrated that the unit cell volume of $Na_3Sc_2(PO_4)_3$ increased monotonously upon heating from 27 to 350 $^oC$ at ambient pressure[13]. According to the inverse-square law, intensity of the radiation is inversely proportional to the square of the distance from the source[14]. On the one hand, intensity of the incident light on phosphor sample is expected to increase with a diminishing distance between the phosphor sample and the excitation source which as a consequence enhance emission intensity of the phosphor[3,4]. On the other hand, the intensity of light out of the phosphor received by the detector increase more considerably due to joint effects of an enhanced emission and a diminished distance between phosphor and detector. Recent work by Song el al. demonstrated that the intensity measurement of WLED phosphors was susceptible to the measuring conditions (especially the sample position) [15]. The amplitude of variation in the optical path lengths for a given spectrofluorimeter is a function of the coefficient of thermal expansion of the phosphor, the amount (or thickness) and packing states of the phosphor powders used for the measurement. It is easily envisaged that the amount of phosphor powders and their packing states for separate measurements or by different technicians should not be exactly the same. It is not unexpected that the magnitude of intensity increase for NSPO: $0.03Eu^{2+}$ phosphor under 370 nm excitation between RT and 165 $^oC$ shown in Fig.4a (25%) and that in Supplementary Fig. 9 (<20%)[1] were inconsistent. This clearly demonstrates that zero-TQ was not an intrinsic property of NSPO:$xEu^{2+}$ phosphors that could be replicated. A reliable calibration and correction for lattice expansion induced optical path length variations in specific temperature-dependent emission and excitation spectra measurements is highly needed.

Table 1 Color coordinates and correlated color temperatures (CCT) of white LEDs fabricated using NSPO:0.03Eu$^{2+}$ phosphor as blue component with yellow-emitting La$_3$Si$_6$N$_{11}$:Ce$^{3+}$ and red-emitting (Sr,Ca)AlSiN$_3$:Eu$^{2+}$ phosphor with UV LED ($\lambda_{max}$ = 365 nm) excitation in the high flux operating current of 100 - 1000 mA

| Current /mA | Color coordinates | | CCT /K* |
| --- | --- | --- | --- |
| | x | y | |
| 100 | 0.2989 | 0.3502 | 7041 |
| 200 | 0.2968 | 0.3531 | 7116 |
| 300 | 0.2958 | 0.3574 | 7121 |
| 400 | 0.2952 | 0.3602 | 7121 |
| 500 | 0.2953 | 0.3617 | 7101 |
| 600 | 0.2957 | 0.3625 | 7075 |
| 700 | 0.2963 | 0.3632 | 7040 |
| 800 | 0.2970 | 0.3633 | 7006 |
| 900 | 0.2975 | 0.3639 | 6978 |
| 1000 | 0.2982 | 0.3639 | 6945 |

*:CCT values are calculated from the CIE color coordinates x and y given in Supplementary Table 5[1] according to McCamy's formula [2]:

$$T (K) = -449n^3 + 3525n^2 - 6823.3 n + 5520.33 \quad (1)$$

*where n= (x-0.3320)/(y-0.1858)*

**Competing financial interests**

The author declares no competing financial interests.

**Supplementary Information**

There were some questionable points in optical characterization of NSPO: xEu$^{2+}$ phosphors. The consistency and reliability of some results are inadequate.

1) The temperature-dependent emission spectra of phosphor NSPO: 0.03Eu$^{2+}$ under 370 nm excitation (Fig.4a and Supplementary Fig. 5) indicated that a maximum of 25% increase in emission intensity in the β-phase compared to that of α-phase at room temperature was observed[1]. However, the magnitude of increase in intensity for the same measurement of the same phosphor presented in Supplementary Fig.9 was less than 20% (yellow solid square)[1]. This evidently indicates that zero-TQ was not an intrinsic property of the phosphor, since the two measurements of temperature-dependent emission by the authors themselves gave inconsistent results.

2) Supplementary Table 7 indicated that the fraction of light absorption by the phosphor remained almost unchanged with increasing temperature irrespective of Eu$^{2+}$ concentration. Nevertheless, Supplementary Fig.3 showed that excitation intensity in the wavelength range of 300–400 nm steadily increased with increasing temperature for all three phosphors[1]. It is well-known that the processes involved in UV-visible region absorption and excitation generally are the same: the electronic transitions from the ground state to the first and second excited states. For excitation to occur, absorption must happen, and if absorption takes place, the system must be excited. In solid crystals like phosphors, the shapes of absorption/excitation bands are determined by the density of states of the initial and final states of electronic states and lattice vibrations. Therefore, absorption and excitation are mutual processes and the two spectra have generally similar shape[2]. It is difficult to understand that temperature dependence of the absorption and excitation spectra of NSPO:xEu$^{2+}$ phosphors differed so considerably.

3) Supplementary Table 7 indicated that the fraction of light absorption by NSPO:0.03Eu$^{2+}$ at temperature range 25–175°C was approximately 0.59[1]. However, it was only 0.43[1] for NSPO:0.07Eu$^{2+}$. According to the authors, Eu$^{2+}$ is the only absorbing and emitting centre of NSPO:Eu$^{2+}$ at room temperature (α-phase)[1]. It is unreasonable that the doping concentration of Eu$^{2+}$ increased from 0.03 to 0.07, whereas the fraction of light absorbed by Eu$^{2+}$ decreased by 27%.

4) Figures 2c & 2d showed that both the intensity and FWHM of Eu$^{2+}$ emission in NSPO:0.07Eu$^{2+}$ under 370 nm excitation remained unchanged upon increasing temperature from 25 to 200 °C. Nevertheless, the temperature-dependent excitation spectra of NSPO:0.07Eu$^{2+}$ (Supplementary Fig. 3 (c) ) indicated that the excitation intensity around 370 nm of NSPO:0.07Eu$^{2+}$ phosphor increased with increasing temperature. It is quite striking that temperature dependence of the excitation and emission spectra of the same phosphor showed such a considerable difference.

5) It is well-known that the 5d-4f luminescence of Eu$^{2+}$ is highly sensitive to its coordinating environment. A slight change in size or symmetry of the coordinating polyhedra of Eu$^{2+}$ often leads to a considerable change of the emission band of the phosphor[3]. The in-situ XRD and Raman spectra showed that the structure of the phosphor host changed from monoclinic α phase to hexagonal β and ϒ phases sequentially during heating[1], indicating that the coordinating environment around Eu$^{2+}$ had changed markedly. It is expected that emission band (peak position $\lambda_{em}$ or the full width at half maximum (FWHM) of Eu$^{2+}$ in NSPO:xEu$^{2+}$ should change with increasing temperature as well. It is difficult to understand that the emission band ($\lambda_{em}$,

FWHM and intensity) of NSPO:0.07Eu$^{2+}$ remained unchanged upon increasing temperature (Fig.2c)[1]. A similar measurement for Na$_3$Sc$_2$(PO$_4$)$_3$ :0.03Eu$^{2+}$ by Wang et al. published in 2016 showed that λ$_{em}$ of the phosphor gradually blue-shifted, accompanied by a gradual peak broadening and drop in intensity with increasing temperature from 25 to 250°C (Supplementary Fig.2) [4].This evidently proves that high thermal stability or zero-TQ is not intrinsic property of NSPO:xEu$^{2+}$ phosphors that could be replicated.